\documentclass[graybox,natbib]{svmult}
\usepackage{mathptmx}
\usepackage{helvet}
\usepackage{courier}
\usepackage{type1cm}
\usepackage{makeidx}
\usepackage{graphicx}
\usepackage{url}
\usepackage{multicol}
\usepackage{chapterbib}
\usepackage[bottom]{footmisc}
\usepackage{arydshln,booktabs,bm} 
\usepackage{array}
\usepackage{dcolumn,warpcol} 
\usepackage{natbib}
\usepackage{here} 
\RequirePackage{amsmath,amssymb}
\usepackage{flafter}
\usepackage{latexsym}

\makeindex
\newcommand{\ccolhd}[1]{\multicolumn{1}{c}{#1}}
\newcommand{\dep}{\pitchfork}
\newcommand{\txt}{\textstyle}
\newcommand{\Greg}{$G^N_{\mathrm{reg}}\,$}

\newcommand{\ci}{\mbox{\protect{ $ \perp \hspace{-2.3ex}
\perp$ }}}

\newcommand{\n}[0]{\hspace*{.35em}}

\newcommand{\nn}[0]{\hspace*{.7em}}

\newcommand{\fourl}[0]{\hspace*{1.4em}}

\newcommand{\node}{\mbox {\LARGE
{$\mbox{$\circ$}$}}}

\newcommand{\snode}{\mbox {\large
{$\mbox{$\circ$}$}}}

\newcommand{\margn}{\mbox {\raisebox{-.1 ex}{\margnn}}}
\newcommand{\margnn}{\mbox {\large
{$\not \: \not $}}$\node $}

\newcommand{\ful}{\mbox{$\, \frac{ \nn \nn \;}{ \nn \nn
}$}}

\newcommand{\fla}{\mbox{$\hspace{.05em} \prec
\!\!\!\!\!\frac{\nn \nn}{\nn}$}}

\newcommand{\fra}{\mbox{$\hspace{.05em} \frac{\nn
\nn}{\nn
}\!\!\!\!\! \succ \! \hspace{.25ex}$}}

\newcommand{\dal}{\mbox{$  \frac{\n}{\n}
\frac{\; \,}{\;}  \frac{\n}{\n}$}}

\graphicspath{{pics/}} 

\begin{document}

\title*{Concepts and a case study for a flexible class of graphical Markov models}
\author{Nanny Wermuth and D. R. Cox (2013).  In:
{\bf Robustness and complex data structures}. Festschrift in honour of Ursula Gather.
Becker, C., Fried, R. \& Kuhnt, S. (eds.) Springer, Heidelberg, 331--350.\\
}
\institute{Nanny Wermuth, Department of Mathematical Sciences, Chalmers University of Technology, Gothenburg, Sweden and International Agency of Research on Cancer, Lyon, France; wermuth@chalmers.se
and D.R. Cox, Nuffield College, Oxford, UK; david.cox@nuffield.ox.ac.uk}
\vspace{-12cm}
\maketitle
\abstract{
With graphical Markov models, one can investigate complex dependences, summarize some results of statistical analyses with graphs and 
use these graphs  to understand implications of well-fitting models. 
The models have a rich history and form an area that has been intensively studied and developed in recent years. We give  a brief review of the main concepts and describe in more detail a flexible  subclass of models, called traceable regressions. These are sequences of joint response regressions for which  regression graphs permit one to trace and thereby understand pathways of dependence.
We use these methods to reanalyze and interpret data from a prospective study of child development, now known as the `Mannheim Study of Children at Risk'. The two related primary features concern cognitive and motor development, at the age of 4.5  and 8 years of a child. Deficits in these features  form 
a sequence of joint responses. Several possible risks are assessed  at  birth of the child and when the child reached  age 3 months and 2 years.}
\section{Introduction}

To observe and understand relations among several features of individuals or objects is one of the central tasks in many substantive fields of research, including the medical, social, environmental and technological sciences. Statistical models can help considerably with such tasks provided they are both flexible enough to apply to a wide variety of different types of situation and  precise enough to guide us in thinking about possible alternative relationships. This
requires in particular  joint responses, which contain continuous random variables, discrete random variables or both types, in addition to only single responses.

Causal inquiries, the search for causes and their likely consequences, motivate much  empirical research. They  rely on appropriate representations of relevant pathways of dependence as they develop over time, often called data generating processes. Causes   which  start  pathways with adverse  consequences may be called  risk factors or risks. Knowing relevant pathways  offers in principle the opportunity to intervene,  aiming to stop the accumulation of some of the risks, and thereby to prevent or at least alleviate their negative consequences.

Properties of persons or objects and features, such as attitudes or behavior of individuals, which can vary for the units or individuals under study, form the variables that are represented in statistical models. A relationship is called a strong positive dependence if knowing one feature makes it much more likely that the other feature is present as well. If, however, prediction of a feature cannot be improved by knowing the other, then the relation of the two is called an independence. Whenever such relations only hold under certain conditions, then they are qualified to be conditional dependences or independences.

Graphs, with nodes representing variables and edges indicating dependences,  serve several purposes. These include to incorporate available knowledge at the planning stage of an empirical study, to summarize aspects important for interpretation after detailed statistical analyses and to predict, when possible, effects of interventions, of alternative analyses of a given set of data  or of changes compared to  results  from  other studies with an identical  core set of variables. 

Corresponding statistical models are called  graphical Markov models. Their  graphs are simple when they have at most one edge for any variable pair even though there may be different types of edge. The graphs can represent different aspects of  pathways, such as the conditional independence structure, the set of all independence statements implied by a graph, or they indicate which variables are needed to generate  joint distributions. In the latter case, the graph represents a research hypothesis on  variables that make an important contribution.
Theoretical and computational work has progressed strongly  during the last few years. 

In the following, we give first  some preliminary considerations. Then we describe some of the history of graphical Markov models and the main  features of their most flexible subclass, called traceable regressions.  We illustrate some of the insights to be gained  with sequences of joint regressions, that turn out to be traceable in a prospective study of child development, now known as  the  Mannheim Study of Children at Risk.

\section{Several preliminary considerations}

Graphical Markov models are of interest in different contexts.  In the present paper, we stress data analysis and interpretation. From this perspective, a number of considerations arise. In a given study, we have objects or individuals, here children, and their appropriate selection into the study is important. Each individual has properties or features, represented as variables in  statistical models.

A first important consideration is that for any two variables, either one is a possible  outcome to the other,  regarded as possibly explanatory, or the two variables are to be treated  as of equal standing. Usually,  an outcome or response refers to a later time period than a possibly explanatory feature. In contrast, an equal standing of two or more features is appropriate when they refer to the same time period or all of them are likely to be simultaneously affected by  an intervention.

On the basis of this, we typically organize the variables for planned statistical analyses into a series of blocks, often corresponding to a time ordering. All relations  between variables within a same block are undirected, whereas those between variables in different blocks are directed in the way described.

An edge between two nodes in the graph, representing a statistical dependence between two variables,  may thus be of at least two types. To represent a statistical dependence of an outcome on an explanatory feature, we use a directed edge with an arrow pointing to the outcome from the explanatory feature.  For relations between features of equal standing,  we use undirected edges. 

In fact, it turns out to be useful to have two types of undirected edge. A dashed line is used to  represent the dependence  between two outcomes or responses given variables in their past. By contrast,  a full line in the "block of variables describing the background or context of the study and early features of the individuals under study,   represents a conditional dependence given all remaining background variables.

From one viewpoint, the role of the graphical representation is to specify statistical independences that can be used to simplify understanding.   From a complementary perspective, often  the more immediately valuable, the purpose is to show those strong dependences that will  be the base for interpreting pathways of dependence.

\section{Some history of graphical Markov models}

The development of graphical Markov models started with undirected, full line  graphs; see  Wermuth (1976), Darroch,  Lauritzen and Speed (1980). The results   built, for discrete random variables, on  the  log-linear  models studied by  Birch (1963), Goodman (1970), Bishop, Fienberg 
and Holland  (1975), and for Gaussian variables,  on the covariance selection models by Dempster (1972). Shortly later, the models were extended
to acyclic  directed graph models for Gaussian and for discrete random variables; see Wermuth (1980), Wermuth and Lauritzen (1983).   
With the new model classes, results from  the beginning of the 20th century  by geneticist  Sewall Wright and by  probabilist  Andrej Markov were combined and extended. 

These generalizations differ from those achieved with  structural equations that were  studied  intensively in the 1950's  within econometrics; see 
for instance Bollen (1989). Structural equation models extend sequences of linear, multiple regression equations by permitting explicitly  endogenous  responses. These have residuals that are correlated with some or all of the regressors. For such endogenous responses, equation parameters need not  measure  conditional dependences, missing edges in graphs of structural equations need not correspond to any independence statement and no simple local modelling may be feasible.
 This contrasts with traceable regressions; see Section 4.1.

 Wright had used directed acyclic graphs, that is graphs with only directed edges and no  variables of equal standing,  to represent  linear generating processes. He developed `path analysis' to judge whether such processes were well compatible with his data.  Path analyses were recognized by Tukey  (1954) to be fully ordered, also called `recursive', sequences of linear multiple regressions in standardized variables.  
 
 With his approach, Wright was far ahead of his time, since, for example,  formal statistical   tests of goodness of fit were developed much later; see Wilks (1938). Conditions  under which directed acyclic graphs represent independence structures for almost arbitrary types of random variables  were studied  later still; see  Pearl (1988), Studen\'y (2005).

 One main objective of traceable regressions  is to uncover graphical representations that lead to an understanding of  data generating processes.  These are not  restricted to linear relations although they may  include linear processes as special cases. A probabilistic data generating process is
a  recursive  sequence of conditional distributions in which response variables can be vector variables that may contain discrete or continuous components or both types. Each of the conditional distributions specifies both the dependences of  a joint response, $Y_a$ say, on components in an explanatory variable vector, $Y_b$, and the undirected dependences among individual  response component  pairs  of $Y_a$.

Graphical Markov models  generalize  sequences of single responses and  single explanatory variables that have been extensively studied as Markov chains.  Markov had recognized at the beginning of the 20th century   that seemingly complex joint probability distributions may  be radically simplified by using the notion of conditional independence. 

In a Markov chain of random variables $Y_1, \ldots,  Y_d$, the joint distribution is built up by starting with the marginal density  $f_d$ of $Y_d$ and   generating then the conditional density $f_{d-1|d}$. At the next step,  conditional independence of $Y_{d-2}$ from $Y_{d}$ given $Y_{d-1}$ is taken into account,  with
$f_{d-2|d-1, d}=f_{d-2|d-1}$.  One  continues  such that   with $f_{i|i+1, \ldots d}=f_{i|i+1}$, response $Y_i$ is conditionally independent of $Y_{i+2}, \dots, Y_{d}$ given $Y_{i+1}$, written compactly in terms of nodes as $i \ci \{i+2, \ldots,  d\}|\{i+1\}$,  and ends, finally,
with $f_{1|2, \ldots, d}=f_{1|2}$, where $Y_1$  has  just $Y_2$ as  an important, directly explanatory variable. 

The fully directed graph, that captures  such a Markov chain, is a single directed  path of arrows.  For five nodes, $d=5$, and  node set  $N=\{1, 2, 3, 4,5\}$, the graph  is $$1 \fla 2 \fla 3 \fla 4 \fla 5 \,.$$
This graph corresponds to a factorization of the joint density $f_N$ given by
$$ f_N=f_{1|2}f_{2|3}f_{3|4}f_{4|5}f_5.$$
The three defining local independence statements given directly by the above factorization or by the graph are: $1\ci \{3,4,5\}|2$,   $2\ci \{4,5\}|3$ and $3\ci 5|4$. One also says that in such a generating  process, each response  $Y_i$ `remembers  of its past just the  nearest neighbour', the nearest past variable  $Y_{i+1}$.

Directed acyclic graphs are the most direct generalization of Markov chains. They   have a fully  ordered sequence of  single nodes, representing individual response variables for which  conditional densities given their past generate $f_N$. No pairs of variables are on an equal standing.  In contrast to a simple Markov chain, in this more general setting, each response may `remember any subset or all of the variables in its past'. 

 Directed acyclic graphs  are also used for  Bayesian networks where the node set may not only consist of random variables, that correspond to features of observable units, but can represent decisions or parameters. As a framework for understanding possible causes and risk factors, directed acyclic are too limited since they exclude the possibility of an intervention affecting several responses simultaneously.

One early   objective of graphical Markov models was to capture   independence structures  by appropriate  graphs. As mentioned before, an independence structure is  the  set of all independence  statements  implied  by the given graph. Such a structure is  to be   satisfied by  any  family of densities, $f_N$, said to be generated over a given graph.  

In principle, all independence statements that arise  from a given set of defining statements of a  graph, may be derived from basic laws of  probability by using the standard properties satisfied by any probability distribution and possibly some additional ones, as described for regression graphs in Section 4.1; see also Frydenberg (1990) for a discussion of properties needed to combine independence statements
captured by directed acyclic graphs.

The  above Markov chain implies for instance also $$ 1\ci  4|3,   \nn \{1, 2\}\ci \{4,5\}|3, \n {\rm and }  \n 2\ci 4|\{1,3,5\}.$$
 For  many variables,  methods defined for   graphs  simplify   considerably  the task of deciding for a given independence statement whether it is implied  by a graphs. Such methods have been called separation criteria; see Geiger, Verma and Pearl  (1990), Lauritzen et al. (1990) and Marchetti and Wermuth (2009) for different but equivalent separation criteria for directed acyclic graphs.

For ordered sequences of vector variables, permitting joint instead of only single responses, the graphs are  directed acyclic in blocks of vector variables. These blocks are sometimes called  the `chain elements'  of the corresponding  `chain graphs'.  
  Four different types of such graphs for discrete variables have been classified and studied  by Drton (2009).
  He proves that two types of chain graph have the desirable property of defining  always curved exponential families for discrete distributions; see for instance  Cox (2006) for the latter concept. 
  
  This property holds for  the `LWF-chain graphs'
of  Lauritzen and Wermuth (1989) and Frydenberg (1990), and for the graphs of Cox and Wermuth (1993, 1996) that  have more recently  been slightly extended  and studied as `regression graphs'; see Wermuth and Sadeghi (2012), Sadeghi and Marchetti (2012). With the added feature that each edge in the graph corresponds to dependence that is substantial in a given context, they become `traceable regressions'; see  Wermuth  (2012). 

 Most books by statisticians on graphical Markov models focus  on  undirected graphs and on  LWF-chain graphs; see  H\o jsgaard, Edwards and Lauritzen (2012), Edwards (2000),  Lauritzen (1996), Whittaker (1990). In this class of graphical Markov models, each dependence between  a  response  and  a  variable  in its past  is considered to be conditional also on all other  components within the same joint response.

   Main  distinguishing features between different types of chain graph are the conditioning sets  for the  independences, associated with the missing edges, and for the  edges present in the graph.
For  regression graphs, conditioning sets are always excluding other components of a given response vector, and
 criteria, to read off the graph all implied independences,   do not change when the last chain element contains an undirected, full-line  graph.  
 It is in this  general form, in which  we introduce this class of models here.  The separation criteria  for these models are generalized versions of the criteria that apply to  directed acyclic graphs. 

\begin{figure} [H]
\centering
\includegraphics[scale=.48]{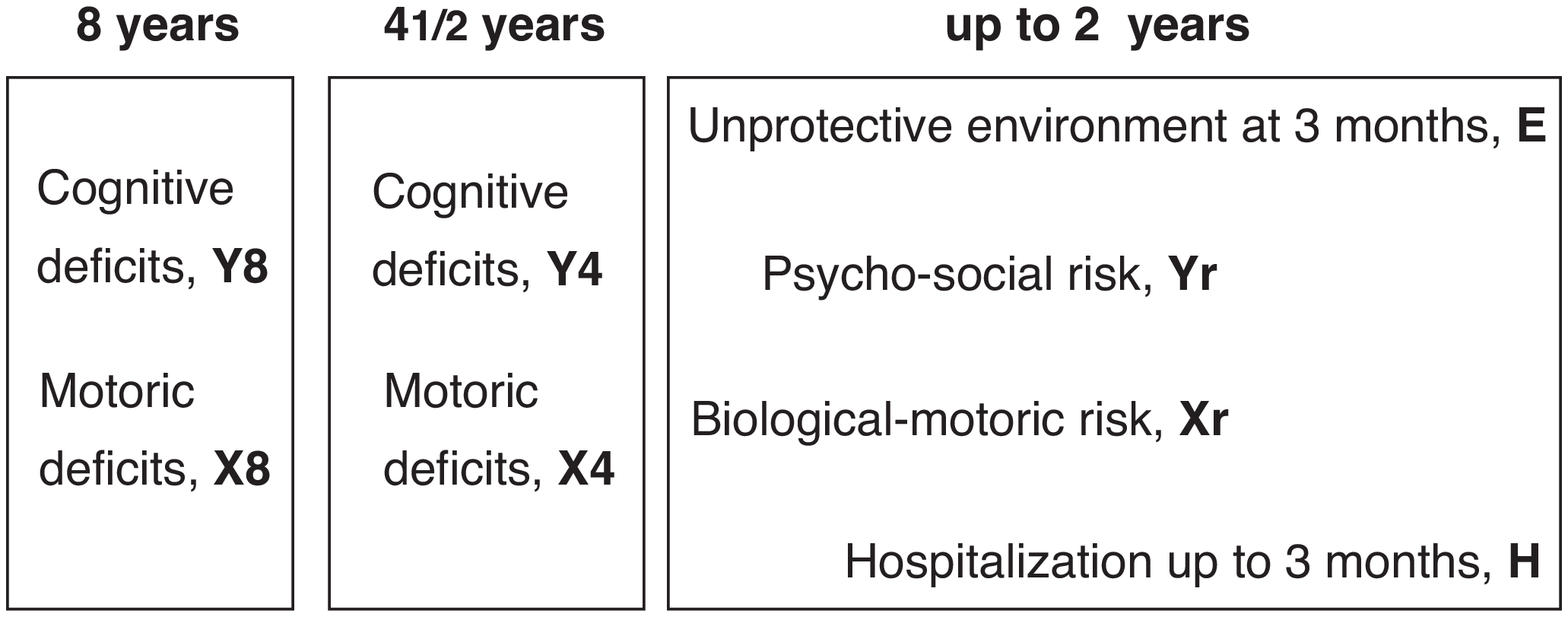} \caption{Ordering of the variables given by time; the joint responses of primary interest are $Y_8, X_8$, those of secondary interest are $Y_4, X_4$,   the four context variables are risks known up to age 2.}
  \label{fig1}   \end{figure}

Figure \ref{fig1} shows two sets of joint responses and a set of background variables, ordered by time.   The two related joint responses concern aspects of 
cognitive and motor development at age 8 years (abbreviated by $Y_8, X_8$, respectively) and at age 4.5 years ($Y_4, X_4$). 
There are two risks, measured  up to 2 years,  $Y_r, X_r$, where $Y_r$ is regarded as  a main risk for cognitive  development and $X_r$ as a main risk for  motor development.  Two more potential risks are available already at age 3 months of the  child. Detailed definitions of  the variables, a description of the study design and of further statistical results are given in Laucht, Esser and Schmidt (1997) and  summarized  in  Wermuth and Laucht (2012). 

 \section{Sequences of regressions and their regression graphs}
\begin{figure} [H]
\centering
\includegraphics[scale=.45]{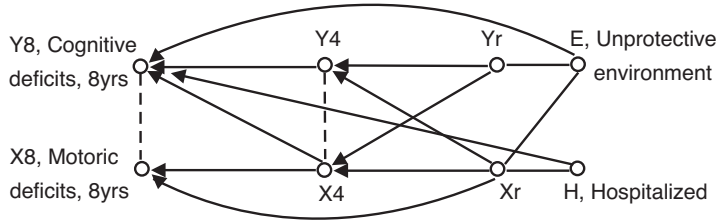} \caption{A well-fitting regression graph for data of the child development  study;
arrows pointing from regressors in the past to a response in the future; dashed lines for dependent responses given their past; full lines for dependent early risk factors given the remaining background variables.}
 \label{fig2}  \end{figure}

The well-fitting regression graph in Figure \ref{fig2} is for the variables of Figure \ref{fig1}
and for data  of 347 families participating in the Mannheim study from birth of their first child until the child  reached the age of 8 years. The graph results from the statistical analyses  reported in Section 4.2. These are further discussed in Section 4.3.

 The goodness-of-fit of the graph to the given data is assessed by local modeling which include here  linear and nonlinear dependences. The following Table 1 gives a summary in terms of Wilkinson's model notation that is in common use for generalized linear mo\-dels and two  coefficients of determination, $R^2$. There is a good fit for quantitative responses when  the changes from $R^2_{\rm full}$ to $R^2_{\rm sel}$ are small,  that is
 from the regression  of an individual response on all  variables in its past  to  a regression  on  only  a reduced set of selected regressors. 
\begin{table} 
\caption{Fitted equations in Wilkinson's notation
}
\begin{tabular}{l l c c}
\toprule
Response \n &  Selected model & $\nn \nn R^2_{\rm full}$\nn & \nn \nn $R^2_{\rm  sel}$\nn \\
 \midrule \\[-2mm]
$ Y_8:$& $Y_4+X_4^2+E+H$ & 0.67& 0.67\\[1mm]
$X_8:$&$ X_4^2+X_r$ &0.36& 0.36\\[1mm]
$ Y_4:$&$ Y_r+X_r^2$ & 0.25 & 0.25\\[1mm]
$X_4:$ &$Y_r+ X_r^2$ & 0.37& 0.36\\[1mm]
$Y_r:$&$E^2$&0.57&0.56\\[1mm]
$X_r:$ &$ E+H$& 0.35& 0.35\\ [1mm]
\bottomrule
\end{tabular}\n\\
Note that any square term implies that also \\
a main effect is included  
\end{table}\n\\[-10mm]

\subsection{Explanations and definitions}

 In each regression graph, arrows point from the past to the future. An arrow is present, between a response and a variable in its past,  when there is a substantively important  dependence,  that is also  statistically significant,  given all its remaining regressors.  Regressors are recognized in the graph by arrows pointing to a given  response node. 
 
 The undirected dependence between two individual components of a response vector is  indicated here by a dashed line; some authors draw instead a bi-directed edge. Such an edge is present if there is  a substantial dependence between two response components given the past of the considered  joint response. An undirected edge between two context variables is a full line. Such an edge is present when there is a substantial  dependence given the remaining context variables.  An edge is missing, when  for this variable pair no dependence can be detected, of the type just decribed.
 
 The important elements of this representation are  node pairs $i,k$, possibly connected by an edge, and a full set ordering  $g_1< g_2< \dots <g_J$ for the connected components $g_j$ of a regression graph.  The connected components of the graph are uniquely obtained by deleting all arrows from the graph and keeping all nodes and all  undirected edges. In general, several  orderings may be compatible with a given graph since different generating 
 processes may lead to  a same  independence structure.
 
There is  further an ordered  partitioning of the node set into two parts, that is a split of $N$ as $N=(u,v)$, such that
 response node sets  $g_1,\dots$ are in $u$ and background
  node sets $\ldots, g_J$ are in $v$. In Figure  \ref{fig2}, there are two  sets in $u$: $g_{1}=\{Y_8, X_8\}$ and $g_{2}=\{Y_4, X_4\}$. The  subgraph of the background variables is for $v=g_3=\{Y_r, X_r, E, H\}$ and there is only one compatible ordering of the three sets $g_j$.
  
 Within $v$,  the  undirected graph is  commonly called  a concentration graph, reminding us of the parameterization for a  Gaussian distribution, where a concentration, an element in the inverse covariance matrix, is a multiple of 
 the partial correlation given all remaining variables; see Cox and Wermuth (1996), Section 3.4, or Wermuth (1976). 
 
 Within $u$,  the  undirected  graph induced by the set $g_j$ is instead a   conditional covariance graph given the past of $g_j$, the nodes in  $g_{>j}=\{g_{j+1}, \dots, g_J\}$; see  Wermuth, Cox and Marchetti (2009), Wiedenbeck and Wermuth (2010)  for  related estimation tasks.
  Arrows may point from any node in $g_j$ for  $j>1$ to  its future in $g_{<j}=\{g_{1}, \dots, g_{j-1}\}$ but never to its past. Thus within each $g_j$, there are only undirected edges  and all arrows  point  from nodes in $g_j$ to nodes in  $g_{<j}$, where $g_{< 1}=\emptyset$.
  
   With $g_{>J}=\emptyset$, the
 basic factorization of  a  family of densities $f_N$, generated over a regression graph, \Greg,  is 
 \begin{equation} f_N= f_{u|v}f_v  \text{ with  }   f_{u|v}=\txt \prod _{g_j\subseteq  u} f_{g_j|g_{>j}}   \text{ and }  f_{v}=\txt \prod _{g_j \subseteq v} f_{g_j} \, 
\label{factdens}, \end{equation}
and the family satisfies all independence constraints  implied by the graph.

For  $i,k$ a node pair, and $c\subset N\setminus\{i,k\}$, we write  $i \ci k|c$ for
  $Y_i, Y_k$  conditionally independent
 given $Y_c$. In terms of a joint conditional density $f_{ik|c}$,  this is equivalent to the following constraints on conditional densities:
$$  i \ci k|c \iff  (f_{i|kc}=f_{i|c}) \iff f_{ik|c}=( f_{i|c}f_{k|c}). $$

 For every  variable pair
$Y_i, Y_k $ making an important contribution to the generating process of $f_N$, we say it is conditionally dependent  given $Y_c$   for some $c\subset N\setminus\{i,k\}$ 
specified in Definition 1 below and write  $i \dep k|c$.  A regression  graph is said to be edge-minimal if every missing edge in the graph corresponds to a conditional independence statement and every edge present is taken to represent  a dependence; see the following definition.
 \begin{definition} \textit{Defining pairwise dependences of } {\Greg}. An edge-minimal  regression graph specifies  with   $g_1<\dots<g_J$ a generating process for $f_N$ where   the following dependences 
 \begin{eqnarray} \label{pairw}
 i \, \dal \, k: \n  i\pitchfork k | g_{>j} \nn \nn \nn  \n &\,&  \text{for } i, k \text{  response nodes in }  g_j \text{ of } u, \nonumber\\
 i \fla k: \n  i \pitchfork k| g_{>j}\setminus \{k\}  &\,&  \text{for  response node } i \text{ in } g_j  \text{ of  }  u  
 \text{ and  node } k \text{  in } g_{>j}, \nn \nn\\
 i\ful k:  \n   i\pitchfork k| v \setminus\{i,k\}  \nn &\,& \text {for }  i, k \text{ context nodes in }  v ,\nonumber
 \end{eqnarray}
define the   edges present  in \Greg.  The meaning of each  corresponding edge missing  in \Greg results with the dependence sign  $\pitchfork$  replaced by  the independence sign $\ci$.
\end{definition}

By equation   \eqref{pairw}, a unique independence statement is assigned to the missing edge of each uncoupled node pair $i,k$. To combine independence statements implied by  a regression graph, two properties are needed, called composition and intersection; see Sadeghi and Lauritzen (2012).  The properties  are  stated below in Definition 3(1) as a same joint independence implied by the  two  independence statements under bullet points 2 and 3 on the right-hand side.
In their simplest form, the two properties can be illustrated with two  simple 3-node graphs. 

For all trivariate probability distributions, one knows 
$i \ci hk \implies (i\ci h \text{ and } i\ci k)$ as well as $i \ci hk \implies (i\ci h|k \text{ and } i\ci k|h)$. The reverse implications are the composition and the intersection property, respectively. Thus, whenever  node $i$ is isolated from the coupled nodes $h,k$ in a 3-node regression graph, it is to be interpreted as $i\ci hk$ and  this type of subgraph in three nodes $i,h,k$ results, under composition, by removing the  $ih$-arrow and the $ik$-arrow in the following graph on the left and under intersection in the following graph on the right. These small examples show already that the two properties 
are used implicitly in the selection of regressors; the composition property for multivariate regressions and the intersection property  for  directed acyclic graph models. 

{\fourl \fourl \fourl  \includegraphics[scale=.45]{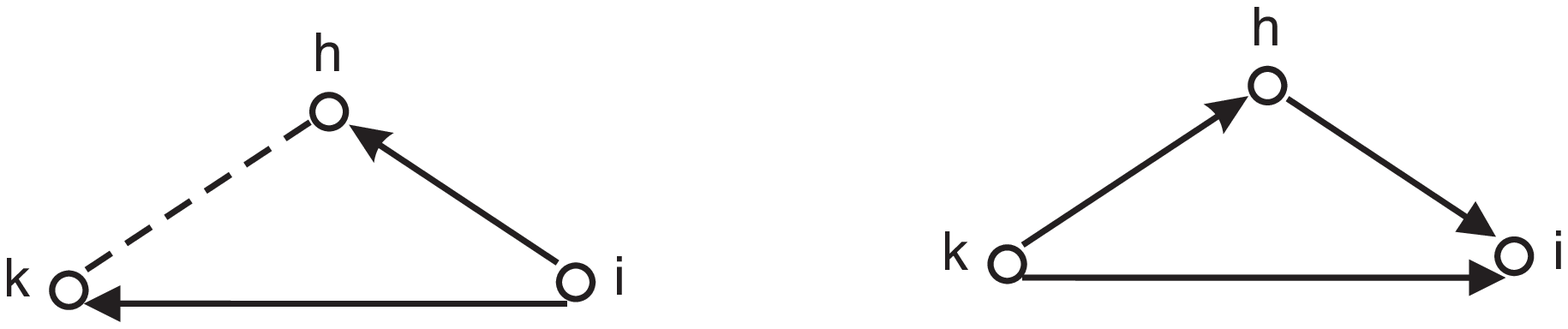}}

For the tracing of  dependences, we need both of these properties but  also the following, called singleton transitivity. It is best explained in terms of 
the {\sf V}s of a regression graph, the subgraphs in 3 nodes having  2 edges.
In a regression graph, there can be  at most 8 different 
{\sf V}-configurations. Such a  {\sf V}  in  three nodes,  $(i, {\rm o}, k)$ say, has uncoupled  endpoints  $i,k$ and  inner node o. 

The {\sf V} configurations  in \Greg are  of two different types. In \Greg,
 the  collision {\sf V}s are:  
$$ i \dal \snode \fla k,  \nn \nn  i  \fra \snode \fla k, \nn \nn i\dal \snode \dal k,$$
and the transmitting {\sf V}s are: 
$$ i \fla \snode \fla k, \nn   i \fla \snode \ful k, \nn   i \ful \snode \ful k,  \nn i \fla \snode \dal  k, \nn 
  i \fla \snode \fra k\,.$$ 
These generalize the 3 different possible {\sf V}s in a  directed-acyclic graph.  For such an edge-minimal graph, the two uncoupled  nodes $i,k$  of a transmitting {\sf V} have either  an important  common-source node (as above on the right) or an important intermediate node (as above on the left), while
the two uncoupled  nodes $i,k$ of a  collision {\sf V} with two arrows pointing to its inner node, have an important, common response.a

Singleton transitivity means
that a unique independence statements is  assigned to the endpoints $i,k$ of each {\sf V} of an edge-minimal graph, either the inner node o is included or excluded in
every independence statement implied by the graph for $i,k$. For the strange parametrisation under which
singleton transitivity is violated in a trivariate discrete family of distributions; see  Wermuth (2012). 

Expressed equivalently, let node pair $i,k$ be uncoupled in  an edge-minimal \Greg and
consider a further  node o and a set $c \subseteq N\setminus \{i,{\rm o}, k\}$. Under singleton transitivity, for both the  independences  $i \ci k|c$  and  $i \ci k|{\rm o} c$ to  hold, one of the constraints ${\rm o} \ci i|c$ or ${\rm o} \ci k|c$ has to be satisfied   as well. Without singleton transitivity, the  path of a {\sf V}
in nodes (i, {\rm o} k) can never induce a dependence for the endpoints $i,k$.

 \begin{definition}  \textit{Dependence-base regression graph}. An edge-minimal \Greg\!, is said to form a dependence base when its defining independences and dependences  are combined by using standard properties of all probability distributions and the three additional properties:
intersection, composition and singleton transiti\-vity. 
 \end{definition}

A dependence base regression graph,  \Greg \!, is edge-inducing by marginalizing over the inner node of a transmitting {\sf V} and by conditioning on the inner node of a collision {\sf V}. This can be expressed more precisely. 
\begin{theorem}{\rm Implications of {\sf V}s in a dependence-base  regression  graph (Wermuth, 2012)}. For each {\sf V}
 in  three nodes,  $(i, {\rm o}, k)$ of  a dependence-base
\Greg, there exists some $c  \subseteq N\setminus\{i, {\rm o}, k\}$,  such that the graph implies ($i \ci k | {\rm o}c$ and  $i\dep k|c$) when it is a transmitting {\sf V}, while it implies   ($i \ci k | c$  and $i\dep k|{\rm o}c$)
when it is a collision {\sf V}.
\end{theorem}

The requirement appears to be elementary, but some densities  or families of densities $f_N$, even when  generated over a dependence base \Greg,
may  have such  peculiar parameterizations that both statements  $i \ci k | {\rm o}c$ and  $i\ci k|c$ can hold even though both node pairs $i$,o and  
o$k$ are coupled by an edge.
 Thus, singleton-transitivity needs to be explicitly carried over to  a generated density. 

We sum up as follows.  For a successful tracing of pathways of dependence in an edge-minimal  regression graph, all three properties are needed: composition, intersection and singleton transitivity. Intersection holds  in all positive distributions and the composition property holds whenever nonlinear and interactive effects also have non-vanishing linear
dependences or main effects. 

Singleton transitivity is  satisfied in  binary distributions;  see Simpson (1951).  More generally, it holds when families of densities are generated  over  \Greg that have a rich enough parametrization, such as the  conditional Gaussian distributions of Lauritzen and Wermuth (1989) that contain discrete and continuous responses.

\begin{definition} \textit{Characterizing properties of traceable regressions.} 
Traceable regression are densities $f_N$ generated over a dependence base  \Greg,  that have for disjoint subsets $a,b,c,d$ of $N$
 \begin{description}
  \item $(1) $ three equivalent decompositions of the same  joint independence 
\begin{itemize}
\item  $b\ci ac|d \iff (b\ci a|cd   \text{ and } b\ci c|d) $
\item    $b\ci ac|d \iff    (b\ci a|d \text{ and } b\ci c|d)$ , \nn 
\item  $b\ci ac|d\iff  (b\ci a|cd \text{ and } b\ci c|ad)$  , \n {\rm and}
\end{itemize}
\item $(2)$  edge-inducing  {\sf V}'s  of  \Greg are  dependence-inducing for $f_N$.  \end{description}
 \end{definition}

One outstanding  feature of  traceable regressions is 
that many of their consequences can be derived by  just using the graph, for instance  when one is marginalizing over some variables in set $M$, and  conditioning  on other variables  in set $C$. In particular,  graphs can be obtained  for node sets $N'=N\setminus\{C,M\}$ which capture  precisely  the independence structure implied by \Greg, the generating graph in the larger node set $N$, for $f_{N'|C}$,  the family of densities of $Y_N'$ given $Y_C$.  

Such graphs are named independence-preserving, when they can be used to derive the independence structure  that would have resulted from the generating graph
by  conditioning on a   larger node set $\{C,c\}$ and marginalizing over the set $\{M,m\}$. Otherwise, such graphs are said to be only independence-predicting. Both types of graph transformations can be based on operators  for binary matrices that represent graphs; see Wermuth,
Wiedenbeck and Cox (2006), Wermuth and Cox (2004).

   From a given generating graph, three  corresponding types of independence-preserving graph result by using the same sets $C,M$.   These are in a subclass of the  much larger class of  MC-graphs of Koster (2002), studied as the ribbon-less graphs by Sadeghi (2012a), or they are the maximal ancestral graphs of  Richardson and Spirtes (2002) or  the summary graphs of Wermuth (2011); see Sadeghi (2012a) for proofs of their Markov equivalence. 

A summary graph shows when a generating conditional dependence,  of $Y_i$ on $Y_k$ say, in $f_N$ remains undistorted 
in $f_{N'|C}$, parametrized in terms of conditional dependences,  and when it be may become  severely distorted; see Wermuth and Cox (2008).
 Some of such distortions can   occur in randomized intervention  studies,  but  they may often be  avoided  by changing  the set $M$ or the set $C$.  
 
 Therefore, these induced graphs  are relevant   for the planning stage of follow-up studies,  designed to replicate some of the results of a given large study by using a subset of the variables, that is after marginalizing over some variables, and/or  by studying a subpopulation, that is after conditioning on another set of variables.
 
For marginalizing alone, that is in the case of $C=\emptyset$, one may apply the following rules for inserting edges repeatedly, keep only one  of several  induced edges of the same type,  and gets often again a regression graph  induced by  $N'=N\setminus M$. In general,  a summary   graph results; see Wermuth (2011). The five transmitting {\sf V}s induce edges by marginalizing over the inner node
$$ i \fla \margn \fla k, \nn   i \fla \margn \ful k, \nn   i \ful \margn \ful k,  \nn i \fla \margn \dal  k, \nn 
  i \fla \margn \fra k\,$$ 
to give, respectively, 
$$ i\fla k,  \fourl \fourl   i\fla k,  \fourl \fourl  i\ful k,  \fourl \fourl  \nn i\dal k,  \fourl \fourl  \nn  i\dal k\,.$$

The induced edges  `remember  the type of  edge at the endpoints of the {\sf V}'  when one takes into account  that each  edge $\snode \dal \snode$ in \Greg can be generated by a larger graph, that contains $\snode\fla \margn \fra \snode$.  Thereby, the independence  structure implied by this graph, for the node  set  excluding the hidden nodes, $\{\margn\}$, is unchanged.

  For any choice of $C,M$ and a given generating graph \Greg, routines in the package `ggm', contained within the computing environment R, help to derive the implications for  $f_{N'|C}$  by computing either one of the  different types of  independence-preserving graph; see Sadeghi and Marchetti (2012). 
  Other routines in  `ggm'  decide whether a given independence-preserving graph is Markov equivalent to another one or to a graph in one of the subfamilies,  such as  a concentration or  a directed acyclic graph; see  Sadeghi (2012b)  for justifications of these procedures.
This helps to contemplate and judge possible alternative interpretations of a  given \Greg.

For two regression graphs, the Markov equivalence criterion is especially simple: the two graphs have to have identical sets of node
pairs with a collision {\sf V}; see Theorem 1 of Wermuth and Sadeghi (2012).
The  result implies that the two sets may contain different ones of the 3 possible collision {\sf V}s.  Also, the two sets of  pairs with a transmitting {\sf V} are then identical, though a given
transmitting {\sf V} in one graph may  correspond  in the other graph to another one  of the 5 transmitting  {\sf V}s that can occur in \Greg\! .

\subsection{Constructing the regression graph via statistical analyses}

As mentioned before, we use here data from the Mannheim  Study of Children at riskt.  The study started in 1986 with a random sample of more than 100 newborns from the general population of children born in the Rhine-Neckar region in Germany. This sample was completed to give equal subsamples, in each of the nine level combinations of two types of adversity, taken to be at levels `no, moderate or high'. In other words, there was heavy oversampling of children at risk. 

The recruiting  of families  stopped with about 40 children of each risk level combination and 362 children in the study. 
All measurements were reported in
standardized form using  the mean and standard deviation of the starting random sample, called here the norm group.  Of  the 362 German-speaking families who
entered the study when their first, single child was born without malformations or any other severe handicap,  347 families participated still when their child reached the age of 8 years.

 Two types of risks were considered, 
one relevant for  cognitive the other for motor development. One main difference to previous analyses  is that we averaged three 
assessments of each type of risk: taken at birth, at 3 months and at two years.   This is justified in both cases by the six  observed  pairwise correlations being all nearly equal.
 The averaged scores, called  `Psycho-social risk up to 2 years', $Y_r$,  and `Biological-motoric risk up to 2 years', $X_r$,
 have smaller variability than the individual components. This points to a more reliable risk assessment and
leads  to  clearly recognizable dependences, to the edges present in Figure \ref{fig2}.

The regression equations may be read off Tables  2 to 7 below. For instance for $Y_8$, there are four regressors and one nonlinear dependence on
$X_4$ with
$${\rm E}_{\rm lin} (Y_8 |{ \rm past \n of \n}Y_8)= 0.03 +  0.78\, Y_8 + (0.07 + 0.10 \,X_4) \,X_4 + 0.11\, E  +   0.12 \,H.$$ 
The test results of  Table  2  imply that the previous measurement of cognitive deficits at age 4 years, $Y_4$ is the most important 
regressor and  that  the next important  dependence  is  nonlinear and on motoric deficits at 4 years, $X_4^2$.

For each individual response component of  the continuous joint responses, the results of linear-least squares fittings are summarized in six tables. In each case,  the response is regressed in the starting model on all the variables in its past. Quadratic or interaction terms are included whenever
there is a priori knowledge or a systematic screening alerts to them; see Cox and Wermuth (1994).

The tables give the estimated constant term and for each variable in the regression, its estimated coefficient
(coeff), the estimated  standard deviation of the coefficient ($s_{\rm coeff})$, as well as the ratio $z_{\rm obs}=$coeff/$s_{\rm coeff}$, often called a studentized value. Each ratio is  compared to the  0.995 quantile of a standard Gaussian random variable $Z$,  for which $\Pr(Z>|2.58|)=0.01$. This relatively strict criterion for excluding variables assures that each edge in the constructed  regression graph corresponds to a dependence  that is considered to be substantively strong in the given context,  in addition to being statistically significant for the given sample size.

 At each backward selection step, the variable
with the smallest observed value $|z_{\rm obs}|$ is deleted from the  regression equation, one at a time, until the threshold is reached so that no more   variables can be excluded. 
The remaining variables are  selected as the regressors of the response. An arrow is added for each of the regressors to the graph containing   just the nodes, arranged in $g_1<g_2< \dots < g_J$.

The 
last column in each table shows the studentized value  $z'_{\rm obs}$, that would be obtained when the variable were included next into
the selected regression equation. Wilkinson's model notation is added in the table to write the selected model in  compact form. For continuous responses, the coefficient of determination is recorded for the starting model, denoted by $R^2_{\rm full}$ and for the reduced model containing the selected regressors, denoted by  $R^2_{\rm sel}$.

A dashed line is added, for a variable pair of a given joint response, when in the regression of one on the other,
there is a significant dependence given their combined  set of the previously selected regressors. 

A full line is added for a variable pair among the background variables,
when in the regression of one on all the remaining background variables, there is a significant dependence of this pair. This exploits that
an undirected edge present in a concentration graph, must also be  be significant in such a regression; see Wermuth (1992).

\begin{table}[H]
\caption{Regression results for  $Y_8$}
\centering
\vspace{2mm}

\setlength{\tabcolsep}{1mm}
\begin{tabular}{l P{-2,2} P{1,2} P{-1,2} c P{-2,2} P{1,2} P{-1,2} c P{-1,2}}
\toprule
\multicolumn{10}{l}{Response: $Y_8$, cognitive deficits at 8 years}\\
\midrule
& \multicolumn{3}{c}{starting model} && \multicolumn{3}{c}{selected} && \ccolhd{excluded}\\
\cmidrule{2-4} \cmidrule{6-8}
explanatory variables & \ccolhd{coeff} & \ccolhd{$s_{\rm coeff}$} & \ccolhd{$z_{\rm obs}$} &&
\ccolhd{coeff} & \ccolhd{$s_{\rm coeff}$} & \ccolhd{$z_{\rm obs}$} && \ccolhd{$z'_{\rm obs}$}\\
\midrule
constant & 0.00  & -\nn&-\nn&& 0.03 &-\nn & -\nn &&-\nn\\
$Y_4$, cognitive deficits, 4.5yrs&  0.78& 0.05 & 15.36 && 0.78 & 0.05 & 15.70 && -\nn\\
$X_4$, motoric deficits, 4.5yrs&  0.05 & 0.04 &  -\nn&& 0.07 &  0.04&  -\nn && -\nn\\
$Y_r$, psycho-social risk, 2yrs&  0.00 & 0.07 & 0.01 && - \nn &  -\nn&  -\nn && -0.13\\
$X_r$, biol.-motoric risk, 2yrs&   0.07 & 0.07 &  1.07 && - \nn &  -\nn&  -\nn && 1.08 \\
$E$, Unprotect. environm., 3mths& 0.10 & 0.06 & 1.81  && 0.12 & 0.04 & 2.62 && -\nn\\
$H$, Hospitalisation up to 3mths& 0.09 & 0.05 & 1.91 && 0.12& 0.04 &  3.00&& -\nn\\

$X_4^2$& 0.09 & 0.01& 6.53 && 0.10 & 0.01 & 7.15 && -\nn\\
\midrule
\multicolumn{10}{l}{$R^2_{\rm full}=0.67$\n\nn Selected model$\n Y_8: Y_4+X_4^2+E+H$ \n \nn$R^2_{\rm sel}=0.67$}\\
\bottomrule \\[-3mm]
\end{tabular}
\label{respY8}
\end{table}

This strategy leads to a well-fitting model, unless one of the excluded variables has a too large contribution 
when it is added alone to a set of selected regressors. Such a variable
would have  to be included as  an additional regressor. 
However, this did not happen for the given set of data.

\begin{table}[H]
\caption{Regression results for  $X_8$}
\centering
\vspace{2mm}

\setlength{\tabcolsep}{1mm}
\begin{tabular}{l P{-2,2} P{1,2} P{-1,2} c P{-2,2} P{1,2} P{-1,2} c P{-1,2}}
\toprule
\multicolumn{10}{l}{Response: $X_8$, motoric deficits at 8 years}\\
\midrule
& \multicolumn{3}{c}{starting model} && \multicolumn{3}{c}{selected} && \ccolhd{excluded}\\
\cmidrule{2-4} \cmidrule{6-8}
explanatory variables & \ccolhd{coeff} & \ccolhd{$s_{\rm coeff}$} & \ccolhd{$z_{\rm obs}$} &&
\ccolhd{coeff} & \ccolhd{$s_{\rm coeff}$} & \ccolhd{$z_{\rm obs}$} && \ccolhd{$z'_{\rm obs}$}\\
\midrule
constant & 0.26  & -\nn&0,26&& -\nn &-\nn & -\nn &&-\nn\\
$Y_4$, cognitive deficits, 4.5yrs&  -0.01& 0.06 & -0.10 && -\nn  & -\nn & -\nn &&0.04 \\
$X_4$, motoric deficits, 4.5yrs&  0.33 & 0.04 & 7.39&& 0.33 &  0.04& -\nn && -\nn\\
$Y_r$, psycho-social risk, 2yrs&  0.01 & 0.08 & 0.19 && - \nn &  -\nn&  -\nn && 0.43\\
$X_r$, biol.-motoric risk, 2yrs&   0.17 & 0.08 &  2.27 && 0.19 &  0.06&  2.97&&  -\nn\\
$E$, Unprotect. environm., 3mths& 0.01 & 0.07 & 0.17  && - \nn & - \nn &  -\nn && 0.44\\
$H$, Hospitalisation up to 3mths& 0.01 & 0.08 & 0.26 && - \nn& - \nn &  -\nn&& 0.26\\

$X_4^2$& 0.18 & 0.23 & 3.41 && 0.05 & 0.02 & 2.89 && -\nn\\
\midrule
\multicolumn{10}{l}{$R^2_{\rm full}=0.36$\n\nn Selected model$\n X_8: X_4^2+X_r$ \n \nn$R^2_{\rm sel}=0.36$}\\
\bottomrule \\[-3mm]
\end{tabular}
\label{respY8}
\end{table}

The tests for the residual dependence of the two response components gives a weak dependence
at age 8 with $z_{\rm obs}=2.4$ but a  strong dependence at age 4.5 with $z_{\rm obs}=7.0$.

 \begin{table}[H]
\caption{Regression results for  $Y_4$}
\centering
\vspace{2mm}
\setlength{\tabcolsep}{1mm}
\begin{tabular}{l P{-2,2} P{1,2} P{-1,2} c P{-2,2} P{1,2} P{-1,2} c P{-1,2}}
\toprule
\multicolumn{10}{l}{Response: $Y_4$, cognitive deficits at 4.5 years}\\
\midrule
& \multicolumn{3}{c}{starting model} && \multicolumn{3}{c}{selected} && \ccolhd{excluded}\\
\cmidrule{2-4} \cmidrule{6-8}
explanatory variables & \ccolhd{coeff} & \ccolhd{$s_{\rm coeff}$} & \ccolhd{$z_{\rm obs}$} &&
\ccolhd{coeff} & \ccolhd{$s_{\rm coeff}$} & \ccolhd{$z_{\rm obs}$} && \ccolhd{$z'_{\rm obs}$}\\
\midrule
constant & -0.29  & -\nn&-\nn&& -0.29 &-\nn & -\nn &&-\nn\\
$Y_r$, psycho-social risk, 2yrs&  0.36 & 0.08 & 4.81 && 0.36&  0.05& 6.77 && -\nn\\
$X_r$, biol.-motoric risk, 2yrs&   0.17 & 0.09 &  -\nn && 0.18 & 0.07&  -\nn &&  -\nn\\
$E$, Unprotect. environm., 3mths& -0.01 & 0.07 & -0.14  && - \nn & - \nn &  -\nn && 0.39\\
$H$, Hospitalisation up to 3mths& 0.14 & 0.04 & 3.36 && - \nn& - \nn &  -\nn&& -0.12\\

$X_r^2$& 0.14 & 0.04 & 3.36 && 0.14 & 0.04& 3.36&& -\nn\\
\midrule
\multicolumn{10}{l}{$R^2_{\rm full}=0.25$,\n\nn Selected model$\n Y_4: Y_r+X_r^2, $ \n \nn$R^2_{\rm sel}=0.25$}\\
\bottomrule \\[-3mm]
\end{tabular}
\label{respY8}
\end{table}

\begin{table}[H]
\caption{Regression results for  $X_4$}
\centering
\vspace{2mm}

\setlength{\tabcolsep}{1mm}
\begin{tabular}{l P{-2,2} P{1,2} P{-1,2} c P{-2,2} P{1,2} P{-1,2} c P{-1,2}}
\toprule
\multicolumn{10}{l}{Response: $X_4$, motoric deficits at 4.5 years}\\
\midrule
& \multicolumn{3}{c}{starting model} && \multicolumn{3}{c}{selected} && \ccolhd{excluded}\\
\cmidrule{2-4} \cmidrule{6-8}
explanatory variables & \ccolhd{coeff} & \ccolhd{$s_{\rm coeff}$} & \ccolhd{$z_{\rm obs}$} &&
\ccolhd{coeff} & \ccolhd{$s_{\rm coeff}$} & \ccolhd{$z_{\rm obs}$} && \ccolhd{$z'_{\rm obs}$}\\
\midrule
constant & -0.47  & -\nn&-\nn&& -0.47 &-\nn & -\nn &&-\nn\\
$Y_r$, psycho-social risk, 2yrs&  0.33 & 0.10 & 3.44 && 0.28 &  0.07&  4.21&&-\nn \\
$X_r$, biol.-motoric risk, 2yrs&   0.62 & 0.11 &  5.50 && 0.50 &  0.09& -\nn&& -\nn\\
$E$, Unprotect. environm., 3mths& -0.06 & 0.08 & -0.66  && - \nn & - \nn &  -\nn && -0.77\\
$H$, Hospitalisation up to 3mths& -0.13 & 0.07 & -1.83 && - \nn& - \nn &  -\nn&& -1.88\\

$(X_r)^2$& 0.21 & 0.05 & 3.97 && 0.23 & 0.05 & 4.43 && -\nn\\
\midrule
\multicolumn{10}{l}{$R^2_{\rm full}=0.37$\n\nn Selected model$\n X_4: Y_r+ X_r^2$ \n \nn$R^2_{\rm sel}=0.36$}\\
\bottomrule \\[-3mm]
\end{tabular}
\label{respY8}
\end{table}
  
\begin{table}[H]
\caption{Regression results for  $Y_r$}
\setlength{\tabcolsep}{1mm}
\begin{tabular}{l P{-2,2} P{1,2} P{-1,2} c P{-2,2} P{1,2} P{-1,2} c P{-1,2}}
\toprule
\multicolumn{10}{l}{Response: $Y_r$, psycho-social risk up to 2 years}\\
\midrule
& \multicolumn{3}{c}{starting model} && \multicolumn{3}{c}{selected} && \ccolhd{excluded}\\
\cmidrule{2-4} \cmidrule{6-8}
explanatory variables & \ccolhd{coeff} & \ccolhd{$s_{\rm coeff}$} & \ccolhd{$z_{\rm obs}$} &&
\ccolhd{coeff} & \ccolhd{$s_{\rm coeff}$} & \ccolhd{$z_{\rm obs}$} && \ccolhd{$z'_{\rm obs}$}\\
\midrule
constant & -0.20  & -\nn&-\nn&& -0.21 &-\nn & -\nn &&-\nn\\
$X_r$, biol.-motoric risk, 2yrs&   -0.04 & 0.04 &  -0.81 && - \nn &  -\nn&  -\nn &&  -1.51\\
$E$, Unprotect. environm., 3mths& 0.57 & 0.03 &-\nn  && 0.55 & 0.03 &  -\nn && -\nn\\
$H$, Hospitalisation up to 3mths& -0.03 & 0.04 & -0.80 && - \nn& - \nn &  -\nn&& -1.50\\

$E^2$& 0.16 & 0.03 & 6.12 && 0.16 & 0.03 & 6.20 && -\nn\\
\midrule
\multicolumn{10}{l}{$R^2_{\rm full}=0.57$\n\nn Selected model$\n Y_r: E^2$ \n \nn$R^2_{\rm sel}=0.56$}\\
\bottomrule \\[-3mm]
\end{tabular}
\label{respYr}
\end{table}

\begin{table}[H]
\caption{Regression results for  $X_r$}
\setlength{\tabcolsep}{1mm}
\begin{tabular}{l P{-2.2} P{-1.2} P{-2.2} c P{-2.2} P{-1.2} P{-2.2} c P{-1.2}}
\toprule
\multicolumn{10}{l}{Response: $X_r$, biologic-motoric risk up to 2 years}\\
\midrule
& \multicolumn{3}{c}{starting model} && \multicolumn{3}{c}{selected} && \ccolhd{excluded}\\
\cmidrule{2-4} \cmidrule{6-8}
explanatory variables & \ccolhd{coeff} & \ccolhd{$s_{\rm coeff}$} & \ccolhd{$z_{\rm obs}$} &&
\ccolhd{coeff} & \ccolhd{$s_{\rm coeff}$} & \ccolhd{$z_{\rm obs}$} && \ccolhd{$z'_{\rm obs}$}\\
\midrule
constant & 0.25  & -\nn&-\nn&& 0.22 &-\nn & -\nn &&-\nn\\
$Y_r$, psycho-social risk, 2yrs&  -0.05 & 0.07 & -0.81 && - \nn &  -\nn&  -\nn && -1.22\\
$E$, Unprotect. environm., 3mths& 0.17 & 0.06 & 3.04  && 0.12& 0.04 & -\nn&& -\nn\\
$H$, Hospitalisation up to 3mths& 0.48 & 0.04 & 12.30 && 0.48 & 0.04& 12.40&& -\nn\\

$E^2$&-0.04 & 0.03 & -1.09 && -\nn  & -\nn & -\nn && -1.42\\
\midrule
\multicolumn{10}{l}{$R^2_{\rm full}=0.35$\n\nn Selected model$\n X_r: E+H$ \n \nn$R^2_{\rm sel}=0.35$}\\
\bottomrule \\[-3mm]
\end{tabular}
\label{Xr}
\end{table}

A global goodness-of-fit test, with proper estimates under the full model, may depend
on additional distributional assumptions and require iterative fitting procedures. For exclusively linear relations of
a joint Gaussian distribution, such a global test for the joint regressions would be equivalent to the fitting of a corresponding 
structural equation model, given  the unconstrained background variables, and  the global fitting of the  concentration graph model
to the context variables would correspond to estimation and testing for one of Dempster's covariance selection models.

\subsection{Using a well-fitting  graph}

There are direct  and indirect  pathways from risks at three months to cognitive deficits at 8 years. The  exclusively positive conditional dependences along  different paths accumulate to positive  marginal dependences, even for responses connected only indirectly to a risk,
for instance  for $Y_8$ to $Y_r$ or $X_8$ to $E$.

Among the background variables, an unprotective environment for the 3 months-old child, $E$,   is strongly related to the  psycho-social risk up to 2 years, $Y_r$
and  hospitalization up to 3 months, $H$,  to the biological-motoric risk up to 2 years, $X_r$. The weakest but still statistically significant dependence  among these four risks occurs   for 
an unprotective environment, $E$, and the biological-motoric risk, $X_r$. 

\begin{figure} [H]
\centering
\includegraphics[scale=.45]{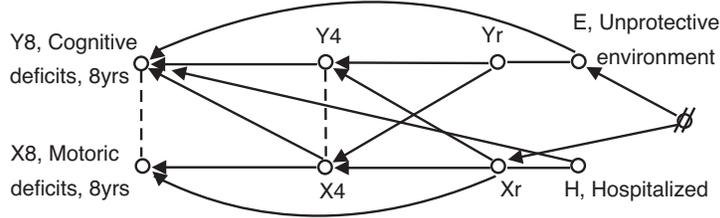} \caption{A graph equivalent to the one of Fig. 2 with one hidden, common explanatory variable}
  \label{fig3}  \end{figure}

Such a dependence taken alone can often best be explained by an underlying
common explanatory variable, here for instance a genetic or a socio-economic risk. This would lead to replacing the full line for $(E, X_r)$ in Figure \ref{fig2} by the common-source
{\sf V}, shown in Figure \ref{fig3}.  The inner node of this {\sf V} is crossed out because it represents a hidden that is unobserved variable.
Hidden nodes represent variables that are unmeasured in a given study but whose relevance and  existence is known or assumed.

\begin{figure} [H]
\centering
\includegraphics[scale=.45]{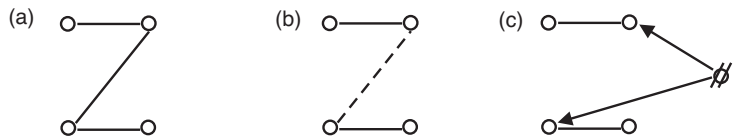} \caption{A hidden variable graph (c) generating two Markov equivalent graphs (a) and (b)}
  \label{fig4}  \end{figure}

Though Figure \ref{fig3}  appears to contain only a small change compared to Figure \ref{fig2}, this change requires a Markov equivalence result for a larger class than regression  graphs, as available
 for the ribbon-less graphs of Sadeghi (2012a), since a path of the type $i\ful {\rm o} \fla k$ does not occur in a regression graph. Given  these results, it follows that graphs Figure \ref{fig4}(a) and (b) are Markov equivalent 
and that the structure of graph \ref{fig4}(b) can be generated by the larger graph \ref{fig4}(c) that includes a common, but hidden regressor  node for the two inner nodes of the path.

To better understand the distinguishing features of the pathways of dependence  in Figure \ref{fig2}  leading to  the joint responses of main interest at  age 8, we generate the implied regressions graphs when the assessments at age 8 and at 4.5 years are available for only one of the 
two aspects. In that case one has ignored,  that is marginalized over, the assessments of the other aspect at age 8 and 4.5.

\begin{figure} [H]
\centering
\includegraphics[scale=.47]{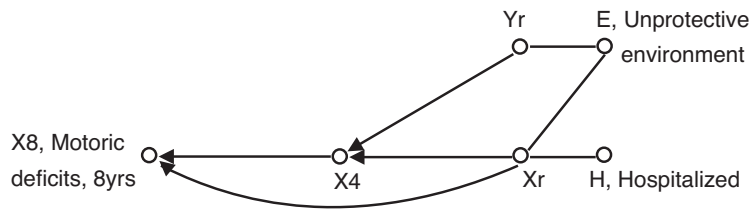} \caption{The regression graph induced by ignoring $Y_8$ and $Y_4$ in Figure \ref{fig2}; $M=\{Y_8,Y_4\}$, $C=\emptyset$}
  \label{fig5}  \end{figure}

The resulting graph, for $Y_8$ and $Y_4$ ignored, happens to coincide with  the subgraph induced by the remaining, selected six nodes in Figure \ref{fig1}, 
as shown in Figure \ref{fig5}. Such an induced graph has the selected nodes and as edges all those present  among them in the starting graph and no more.       
 The graph of Figure
\ref{fig5}  implies that possible psycho-social risks of a child up to age 2, $Y_r$,  do  not contribute directly to predicting motoric deficits at school-age, $X_8$,  also  when the more recent information on cognitive deficits is not available.

\begin{figure} [H]
\centering
\includegraphics[scale=.47]{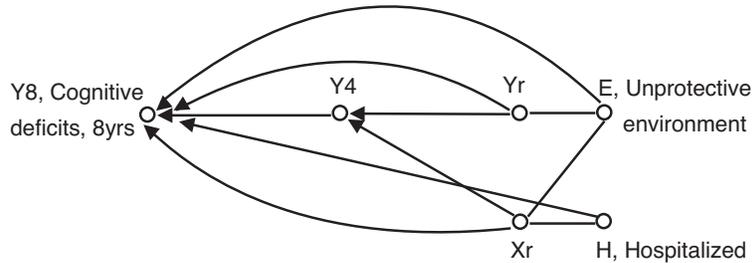} \caption{The regression graph induced by ignoring $X_8$ and $X_4$ in Figure \ref{fig2}; $M=\{X_8,X_4\}$, $C=\emptyset$}
  \label{fig6}  \end{figure}

By contrast, the regression graph  in Figure \ref{fig6} that results after  ignoring $X_8$ and $X_4$,   shows two additional arrows 
compared to the subgraph induced in Figure \ref{fig2} by $Y_8,Y_4, Y_r, X_r, E,H$. 
  The induced arrows are for $ (Y_8, Y_r)$ and for
$(Y_8, X_r)$. The graph suggests that cognitive deficits at school-age, $Y_8$, are directly  dependent on all of the remaining variables when the more recent  information
on the motoric risks are unrecorded. There are direct and indirect pathways from $H$ and from $E$ to $Y_8$. They  involve nonlinear
dependences of cognitive deficits on previous motoric deficits or risks. These are recognized in the fitted equations but not directly in the graph alone.

What the graph also cannot show is that with $X_8,X_4$ unrecorded, the early risks, $Y_r, H$ are  less important as predictors when   $Y_4, X_r, X_r^2, E$ are available as regressors of $Y_8$. This effect is due to the strong partial dependences of $Y_r, E^2$ given $E, X_r, H$ and of $X_r,H$ given $E,E^2, Y_r$. Such implications, due to the special parametric constellations are not reflected in the graph alone.

   Many more conclusions may be  drawn by using just  graphs like in Figures \ref{fig2} to \ref{fig6}.  The substantive research questions and the special  conditions of a given study are important; for some different types of  study analyzed with graphical Markov models see, for instance,  Klein, Keiding and Kreiner (1995), Gather, Imhoff and Fried (2002),  Hardt et al (2004), Wermuth, Marchetti and Byrnes (2012).  
  
  One major attraction of  sequences  of regressions in joint responses is that they may model  longitudinal data from observational as well as from intervention studies. For instance,  with fully randomized allocation of persons to a treatment, all  arrows that may point to the treatment in an observational study,  are removed from the regression graph.  This removal reflects such a successful  randomization: independence is assured for the treatment variable of all regressors or background variables, no matter whether they are observed or hidden.

\section{Conclusions}

The paper combines two main themes. One is the notion of traceable regressions. These  are sequences of joint response regressions together
with a set of background variables for which an associated  regression graph not only captures an independence  structure but permits the tracing of pathways of dependence. Study of such structures has both a long history and at the same time is the focus for much current development. 

Joint resposes are needed when causes or risk factors  are expected to affect several responses simultaneously. Such situations occur frequently and cannot be adequately modeled with distributions generated over directed acyclic graph or such a graph with added dashed lines between responses 
and variables in their past to permit  unmeasured confounders or endogenous responses.

A regression graph shows, in particular, conditional independences by missing edges and conditional dependences by edges present.
The independences  simplify  the  underlying data-generating process  and  emphasize the important  dependences via the remaining edges.
The  dependences  form the basis for interpretation, for the planning of or comparison with further studies  and for possible policy action.  Propagation of independencies  is now reasonably well understood. There is scope for complementary further study that  focuses on pathways of dependence.

The second theme concerns specific applications. Among the important issues here are an appropriate definition of population under study, especially  when relatively rare events and conditions  are to be investigated,  appropriate sampling strategies,  and the importance of building an understanding on step-by-step local analyses.
The data of the Mannheim study happen to satisfy all properties needed for tracing pathways of dependence. This permits discussion of the  advantages and limitations for some illustrated path tracings.
  
  In the near future,  more results on estimation and goodness of fit tests are to be expected, for instance  by extending the fitting procedures for regression graph models of  Marchetti and Lupparelli  (2010) to mixtures of discrete and continuous variables,  more results on the identification of models that include hidden variables such as those by Stanghellini and Vantaggi (2012) and those by Foygel, Draisma and Drton (2012),  and  further evaluations of properties of different  types of parameters; see Xie, Ma and Geng (2008) for an excellent starting discussion.\\
  
  \noindent{\large \bf Acknowledgement} The work by Nanny Wermuth reported in this paper was undertaken during her tenure of a Senior Visiting Scientist Award by the International Agency of Research on Cancer. We thank the referees, Bianca de Stavola and Rhian Daniel for their helpful  comments. We used Matlab  for statistical analyses.

\end{document}